\documentclass[aps,pre,reprint,showpacs,twocolumn]{revtex4}

\usepackage{graphicx}
\usepackage{float}
\usepackage{color}
\usepackage{amsmath}
\usepackage{amsfonts}
\usepackage{amssymb}
\usepackage{amsbsy}
\usepackage{url}
\usepackage{bbold}

\DeclareMathOperator{\Tr}{Tr}

\begin{document}

\title{Phase Transition of the four-dimensional Cross-Polytope Model}

\author{Roman Krcmar}
\affiliation{Institute of Physics, Slovak Academy of Sciences, SK-845 11, Bratislava, Slovakia}
\author{Andrej Gendiar}
\affiliation{Institute of Physics, Slovak Academy of Sciences, SK-845 11, Bratislava, Slovakia}
\author{Peter Rapcan}
\affiliation{Institute of Physics, Slovak Academy of Sciences, SK-845 11, Bratislava, Slovakia}
\author{Tomotoshi Nishino}
\affiliation{Department of Physics, Graduate School of Science, Kobe University, Kobe 657-8501, Japan}

\begin{abstract}

Thermodynamic properties of the four-dimensional cross-polytope model, the 16-cell model, 
which is an example of higher dimensional generalizations of the octahedron model, are 
studied on the square lattice.
By means of the corner transfer matrix renormalization group (CTMRG) method, presence of 
the first-order phase transition is confirmed. 
The latent heat is estimated to be $L_4^{~} = 0.3172$, which is larger than that of the octahedron 
model $L_3^{~} = 0.0516$. 
The result suggests that the latent heat increases with the internal dimension $n$ when the 
higher-dimensional series of the cross-polytope models is considered.

\end{abstract}
\maketitle

\section{Introduction}

There is a variety of spin models that contain vectors of unit length on regular lattices as site 
degrees of freedom. Typical examples are the $n$-vector models, which have the $O( n )$ 
symmetry~\cite{Stanley}. When the internal space dimension $n$ is equal to or larger than $2$, 
the $O( n )$ symmetry is continuous. Thus the $n$-vector models do not exhibit any fixed order 
in finite temperature when $n \ge 2$ if the models are defined on one of the two-dimensional regular 
lattices, as it was proved by Mermin and Wagner~\cite{MW}. The two-vector model is known as 
the classical XY model, which allows the presence of Berezinskii-Kosterlitz-Thouless (BKT) 
transition~\cite{Bere1,Bere2,KT} on two-dimensional lattices. An ordered state can appear when 
discrete nature or perturbation is introduced to the $O( n )$ symmetry. The two-dimensional 
ferromagnetic $q$-state clock models, which are the discrete analogues of the two-vector model, 
exhibit a fixed ordering when temperature is sufficiently low.

Discrete analogues of the three-vector model, which is the classical Heisenberg model, exhibit
characteristic phase transitions on the two-dimensional lattices, and 
their thermodynamic properties depend on the type of 
the spin discretization. The polyhedron models are typical examples, where the site degrees of
freedom are represented by the unit vectors pointing to all vertices of a regular polyhedron. 
For instance, the tetrahedron model, which can be mapped to the four-state Potts model, exhibits
a second-order phase transition with logarithmic corrections~\cite{Wu}. The octahedron model,
which has six-states, exhibits a weak first-order phase transition~\cite{Patrascioiu1,Roman}.
The cubic model with eight states is equivalent to the three independent Ising models. Patrascioiu 
{\it et al.} reported a second-order phase transition for both the dodecahedron model (with 12 states)
and the icosahedron model (with 20 states)~\cite{Patrascioiu1,Patrascioiu2,Patrascioiu3}.
Recently, presence of the second-order phase transition has been confirmed for these two 
models by means of extensive numerical calculations~\cite{Ueda,Ueda2}, where a parallelized 
version of the corner transfer matrix renormalization group (CTMRG) 
method~\cite{Baxter1,Baxter2,Baxter3,Okunishi1,Okunishi2,Okunishi3} was used. 
The universality classes of these 12 and 20 state models have not been fully identified yet. 

There are three families of the discrete analogues of the $n$-vector model that can exist for
arbitrary $n$. The first family consists of simplex models, which include the three-state clock
model ($n = 2$), the tetrahedron model ($n = 3$) and the $5$-cell model ($n = 4$). It is known
that the $q$-cell model ($q = n + 1$) can be mapped onto the $q$-state Potts models~\cite{Wu},
which exhibit the first-order phase transition for $q \ge 5$. The second family consists of
hyper-cubic models, and they include the four-state clock model ($n = 2$), the cube model
($n = 3$), and the $8$-cell model ($n = 4$). The site degrees of freedom are equal to $2^n_{~}$,
and it is straightforward to show that those models are equivalent to a set of $n$ independent
Ising models. Thus, within this family, the phase transitions always remain of the second-order
type. Finally, the third family consists of the cross-polytope models, which include the
octahedron model ($n = 3$) and its higher-dimensional generalizations.
In this article we mostly focus on the
four-dimensional case ($n = 4$), which corresponds to the $16$-cell model, and 
analyze its phase transition. The free energy, internal energy, and spontaneous magnetization
are calculated by the CTMRG method. We observed the first-order phase transition, where the
latent heat is obtained as $L_4^{~} = 0.3172$. Additionally, we also re-examined
the octahedron model, and re-estimated its latent heat to obtain $L_3^{~} = 0.0516$. 
The calculated results suggest that $L_n^{~}$ is an increasing function of the internal
dimension $n$.

The structure of this article is as follows. In the next section, we introduce the
cross-polytope models. Numerical results by the CTMRG method are shown in Sec.~III.
Conclusions are summarized in Sec.~IV. We also discuss the remaining studies for
various discrete $O( n )$ models.

\section{Cross-polytope models}

Let us now introduce the discrete analogues of the $n$-vector models that belong to the cross-polytope 
family. Suppose that there is a vector of the unit length ${\boldsymbol{S}}_i^{~}$ on each lattice point
of the square lattice. 
The index $i$ 
specifies the location 
of the site 
on the square lattice.
The vector can point to the vertices of the cross-polytope. 
For example, if $n = 3$, a vector ${\boldsymbol{S}}_i^{~}$ 
of the octahedron model can be one of the six vectors
\begin{equation}
( \pm 1, 0, 0 ) \, ,  ~ ( 0, \pm 1, 0 ) \, , ~{\rm and}~ ( 0, 0, \pm 1 ) \, .
	\label{8-vecs}
\end{equation}
If $n = 4$, the site vector
${\boldsymbol{S}}_i^{~}$ of the 16-cell model can be 
any one of the eight vectors
\begin{eqnarray}
&& ( \pm 1, 0, 0, 0 ) \, , ~~~~~~~ ( 0, \pm 1, 0, 0 ) \, , ~ \nonumber\\
&& ( 0, 0, \pm 1, 0 ) \, , ~{\rm and}~  ( 0, 0, 0, \pm 1 ) \, .
	\label{16-vecs}
\end{eqnarray}
In general, ${\boldsymbol{S}}_i^{~}$ is an $n$-dimensional vector where only one
component is either $1$ or $-1$, whereas all other components are $0$. 
Thus, there are  $2n$ site degrees of freedom in total. 

In the following, we consider the model whose Hamiltonian is expressed as
\begin{equation}
H = - J \sum_{\langle i, j \rangle}^{~} {\boldsymbol{S}}_i^{~} \cdot {\boldsymbol{S}}_j^{~} \, ,
\label{Hm}
\end{equation}
where $\langle i, j \rangle$ represents the neighboring pairs on the square lattice. 
The inner product ${\boldsymbol{S}}_i^{~} \cdot {\boldsymbol{S}}_j^{~}$ can results in $1$, $0$,
or $-1$. For simplicity, we set the ferromagnetic coupling $J = 1$ 
throughout this article.

Thermodynamic properties of the system are obtained through the partition function 
formally written as
\begin{equation}
Z = \Tr \, e^{- \beta H}_{~} \, ,
\end{equation}
which is the function of $\beta = 1 / k_{\rm B}^{~} T$, where $T$ is the temperature and 
$k_{\rm B}^{~}$ represents the Boltzmann constant. For simplicity, we set $k_{\rm B}^{~} = 1$ 
in the following. We have expressed the configuration sum for all the vectors on the lattice 
using the trace notation.  For the convenience in the numerical calculations, 
we express the system as the interaction round a face (IRF) model, where each 
IRF weight is represented by the Boltzmann factor
\begin{eqnarray}
&& \!\!\!\!\!\!\!\!\!\!\!\!
W( \boldsymbol{S}_i^{~}, \boldsymbol{S}_j^{~}, \boldsymbol{S}_k^{~}, \boldsymbol{S}_{\ell}^{~} ) 
\nonumber\\
&& \!\!\!\!\!\!\!\!\!\!\! = \exp\biggl[ \frac{\beta J}{2} \! \left(
\boldsymbol{S}_i^{~}\cdot\boldsymbol{S}_j^{~} + 
\boldsymbol{S}_j^{~}\cdot\boldsymbol{S}_k^{~} + 
\boldsymbol{S}_k^{~}\cdot\boldsymbol{S}_{\ell}^{~} + 
\boldsymbol{S}_{\ell}^{~}\cdot\boldsymbol{S}_i^{~} 
\right) \biggr] ,
\end{eqnarray}
where the vectors $\boldsymbol{S}_i^{~}$, $\boldsymbol{S}_j^{~}$, $\boldsymbol{S}_k^{~}$, and 
$\boldsymbol{S}_{\ell}^{~}$ are located on the corners of a square-shaped unit cell. 

A variety of the thermodynamic functions can be obtained from the
free energy 
\begin{equation}
F = - k_{\rm B}^{~} \, T \, \ln \, Z \, .
\end{equation}
Alternatively, one-point functions can be directly calculated from the thermal average with
respect to the Boltzmann factor $e^{- \beta H}_{~}$. An example is the spontaneous magnetization 
per site
\begin{equation}
M( T ) = \langle {\boldsymbol{S}}_i^{~} \rangle = 
\frac{1}{Z} \, {\rm Tr} \, \Bigl[ {\boldsymbol{S}}_i^{~} \cdot \boldsymbol{\sigma}  \,\, e^{- \beta H}_{~} \Bigr] \, ,
	\label{SM}
\end{equation}
which is independent on the location $i$ in the thermodynamic limit, 
and is finite in the low-temperature ordered state. 
The unit vector $\boldsymbol{\sigma}$ represents the direction of the ordering. Another
example is the bond energy 
\begin{equation}
U( T ) = -J\langle {\boldsymbol{S}}_i^{~} \cdot {\boldsymbol{S}}_j^{~} \rangle 
= 
-\frac{J}{Z} \, {\rm Tr} \, \Bigl[ {\boldsymbol{S}}_i^{~} \cdot {\boldsymbol{S}}_j^{~}
	\,\, e^{- \beta H}_{~} \Bigr] \, ,
	\label{IE}
\end{equation}
where ${\boldsymbol{S}}_i^{~}$ and ${\boldsymbol{S}}_j^{~}$ are the nearest neighbors. 

If the first-order phase transition is present in the discrete $n$-vector model
of the polytope type, the latent heat
\begin{equation}
L_n^{~} 
= \lim\limits_{T \to T_{n}^{+}} U(T) \,\, - \lim\limits_{T \to T_{n}^{-}} U(T)
\label{LH}
\end{equation}
is finite, where $T_{n}^{+}$ and  $T_{n}^{-}$, respectively, denote the limits 
to the transition temperature $T_{n}$ from high- and low-temperature side.

\section{Numerical results}

We consider a set of finite-size systems with the square geometry, 
and denote the size of each system by the linear dimension $N$.
Let us express the corresponding partition function by $Z( N )$.
We calculate $Z( N )$ iteratively by means of the CTMRG method, 
starting from $Z( 3 )$ and increasing the system size $N$ by $2$ in each 
numerical iteration step.
It is possible to choose boundary conditions for the square systems, 
by setting the appropriate conditions for the initial tensors.
Under the fixed boundary condition, all the boundary vectors are kept aligned in 
an identical direction. Under the free boundary condition, the boundary vectors 
can point to arbitrary directions.

It should be noted that the numerically calculated value for $Z( N )$ 
is slightly dependent on
the number of the renormalized block-spin state $m$ in the CTMRG method. 
We denote the approximated value by $Z( N, m )$. 
We keep $m = 300$ states at most, the condition of which enables us to estimate
the latent heat $L_n^{~}$ quantitatively in the large-$m$ limit.  Let us 
introduce the calculated free energy per site
\begin{equation}
f_n^{~}( N, m ) = - \frac{1}{N^2_{~}} \,  k_{\rm B}^{~} T \, \ln \, Z( N, m ) \, .
\end{equation}
In the cases $n = 3$ and $4$ that we examine in the following, 
the convergence of $f_n^{~}( N, m )$ with respect to $N$ 
to the limit $f_n^{~}( \infty, m )$
is fast enough regardless of the 
temperature $T$. Such a rapid convergence suggests that the system is always off-critical. 

\begin{figure}[tb]
\includegraphics[width = 0.45\textwidth]{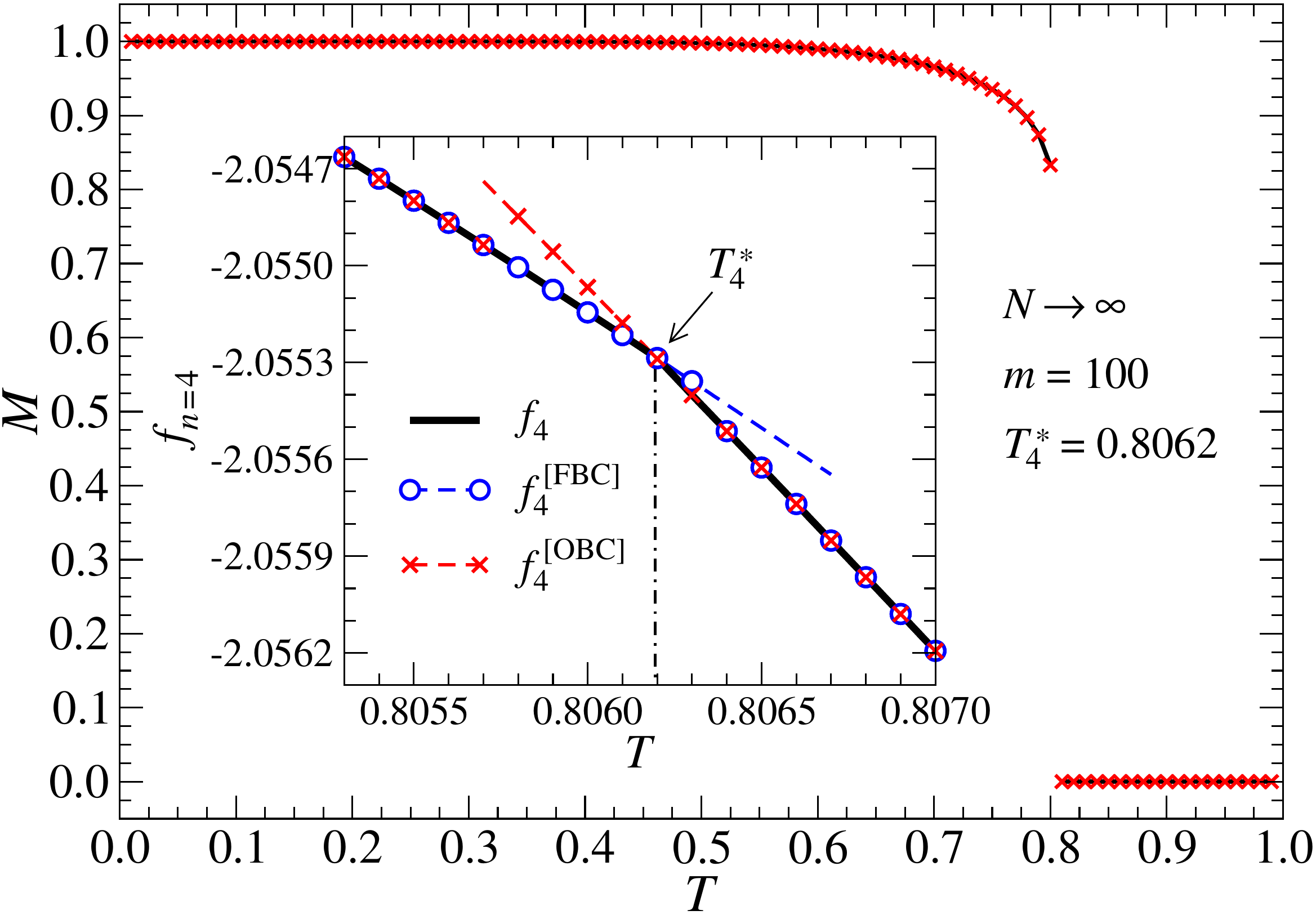}
\caption{Spontaneous magnetization $M$ of the 16-cell model ($n=4$) 
with respect to $T$ in the limit $N \rightarrow \infty$ when $m = 100$.
There is a discontinuity at $T^{*}_{4} = 0.80620$.  
The inset shows the free energy per site under
the fixed boundary conditions $f^{\rm [FBC]}_{4}$ and the open 
ones $f^{\rm [OBC]}_{4}$.}
\label{fig1}
\end{figure}

We first show the calculated result for the case  $n=4$, the 16-cell model. 
Figure~\ref{fig1} shows the spontaneous magnetization 
$M( T )$ in Eq.~\eqref{SM} in the thermodynamic limit $N \rightarrow \infty$, 
calculated under the condition $m = 100$.  There is a discontinuity at the 
temperature $T^*_{4} = 0.80620$. 
(We put `$*$' mark for the calculated cross-over temperature, which can be
dependent on $m$.)
The behavior suggests the presence of the first-order
 phase transition.  In order to get complementary information, 
we observe the effect of boundary conditions on the
free energy per site $f_4^{~}( \infty, m )$. The inset of Fig.~\ref{fig1} shows 
$f^{\rm [FBC]}_{4}( \infty, m )$ under the fixed boundary conditions and 
$f^{\rm [OBC]}_{4}( \infty, m )$ under the open ones. 
As it is shown, there is a crossover at $T^*_{4} = 0.80620$ 
between the ordered state at low temperatures and the disordered one 
at high temperatures.

\begin{figure}[tb]
\includegraphics[width = 0.45 \textwidth]{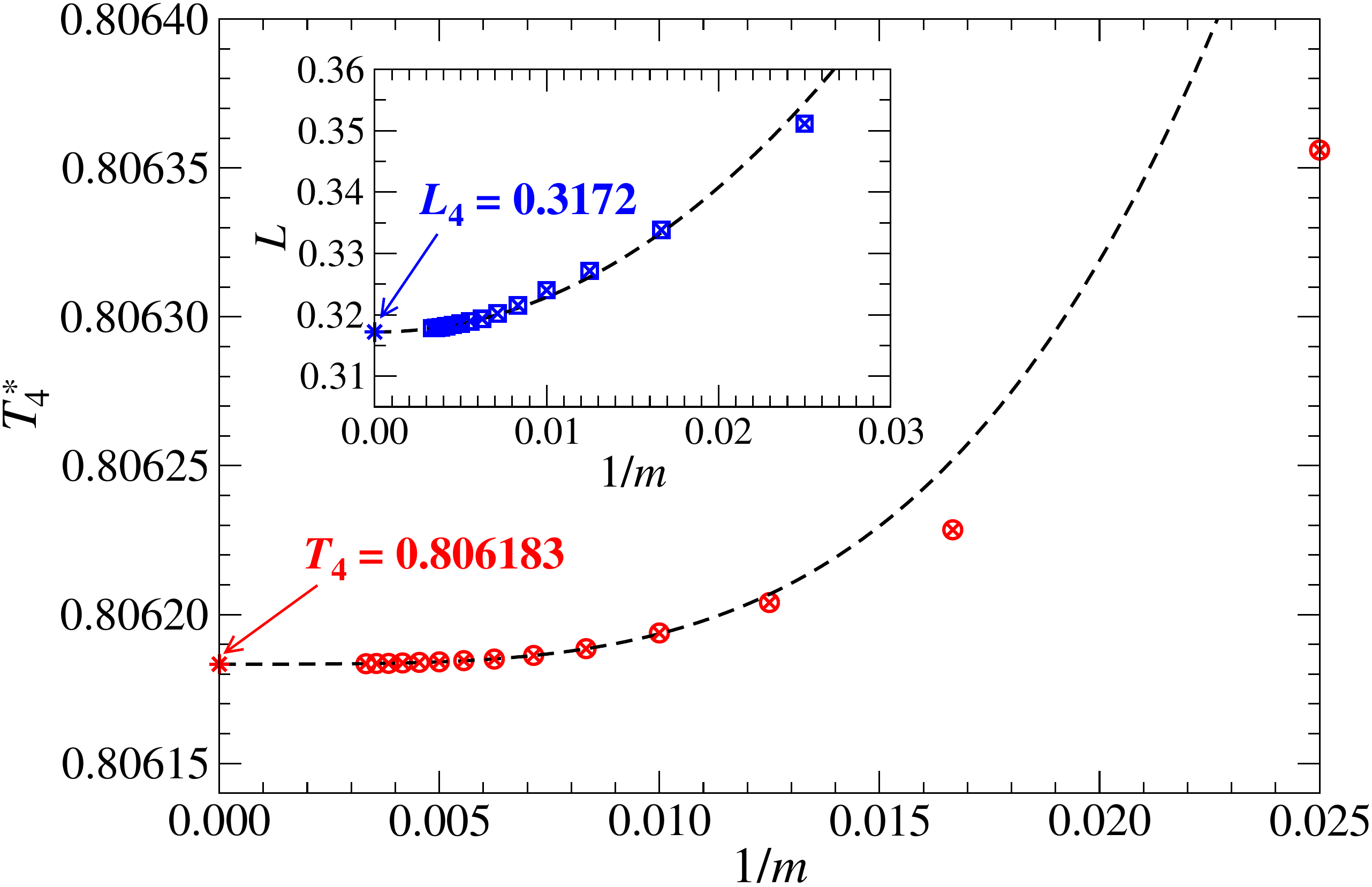}
\caption{Crossover temperature $T^*_{4}$ with respect to $1/m$. 
The dashed line shows the fitting result to the function
$T_4^{~} + c \, e^{a/m}_{~}$, where $T_4^{~} = 0.80618$ is obtained. 
The inset shows the jump in the calculated internal energy per site
at $T^*_{4}$. Taking the limit $m \to \infty$ the latent heat is estimated 
as $L_4^{~} = 0.3172$.}
	\label{fig2}
\end{figure}

Precisely speaking, the crossover temperature $T^*_{4}$ 
slightly depends on $m$, even around $m = 100$.
Figure~\ref{fig2} shows the $m$-dependence of $T^*_{4}$, 
which is almost converged around $m = 200$. Fitting the 
plotted data with the function $T_4^{~} + c \, e^{a/m}_{~}$
within the range $100 \le m \le 300$, we estimate 
the phase-transition temperature $T_4^{~} = 0.806183$. 
The inset of Fig.~\ref{fig2} shows the difference of the internal energy per site
between the ordered and the disordered states at the crossover temperature $T^*_4$. 
Even when $m = 300$, there is a non-negligible $m$-dependence, and, therefore, we perform 
the extrapolation with the use of the fitting function 
$L_{4}^{~} + c \, e^{b/m}_{~}$ within the range $160 \le m \le 300$.
As a result, we estimate the latent heat $L_4^{~} = 0.3172$. 

\begin{figure}[bt]
\includegraphics[width=0.45\textwidth]{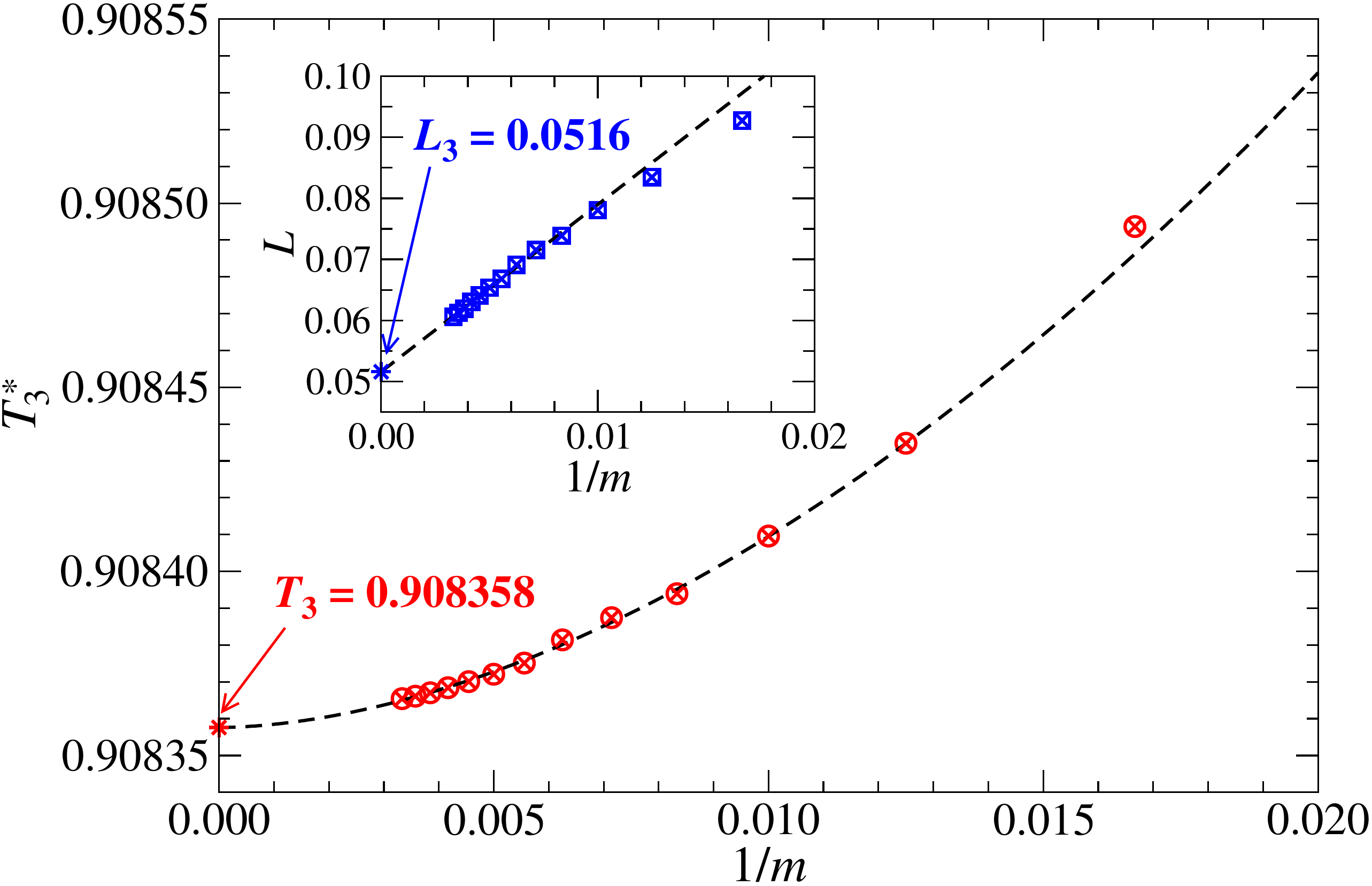}
\caption{Crossover temperature $T^*_{3}$ of the octahedron model ($n = 3$) 
with respect to $1 / m$. Fitting to $T_3^{~} + c \, e^{a/m}_{~}$, we obtain
$T_3^{~} = 0.908358$. The inset shows the jump in the calculated internal
energy versus $1/m$. The linear fitting yields $L_3^{~} = 0.0516$.}
	\label{fig3}
\end{figure}

For comparison, let us focus on the thermodynamic property of the octahedron model
($n = 3$). In our previous study, we obtained the transition temperature
$T_3^{~} = 0.908413$ from the CTMRG calculation under $m = 300$ only, where we 
estimated the latent heat to be $L_3^{~} = 0.073$~\cite{Roman}. In the study 
the $m$-dependence in $T_3^*$ was not carefully examined. 
We thus perform re-estimation of $T_3^{~}$ and $L_3^{~}$.
In the same manner as we have analyzed the 16-cell model,
we calculate $f^{\rm [FBC]}_{3}( \infty, m )$ and $f^{\rm [OBC]}_{3}( \infty, m )$ 
in order to determine their crossover temperature $T^*_{3}$. In Fig.~\ref{fig3}, 
we plot $T^*_{3}$ with respect to $1/m$. Fitting the plotted data to the function 
$T_3^{~} + c \, e^{a/m}_{~}$ within the range $80 \le m \le 300$, we obtain
$T_3^{~} = 0.908358$, which is slightly lower than the value we had previously
reported~\cite{Roman}. The inset of Fig.~\ref{fig3} shows the jump in the internal
energy per site with respect to $1/m$. In this case, the exponential convergence
is not observed. We, therefore, carried out a linear fit to the plotted data within
the range $80 \le m \le 300$. The estimated latent heat, $L_3^{~} = 0.0516$, is
smaller than the value we had reported earlier~\cite{Roman}.
It is worth to mention that the estimated $L_3^{~} = 0.0516$ is of the 
same order as the latent heat
measured in the 5-state Potts model $L \sim 0.0265$~\cite{Wu,Baxter3}.

\begin{figure}[tb]
\includegraphics[width=0.45\textwidth]{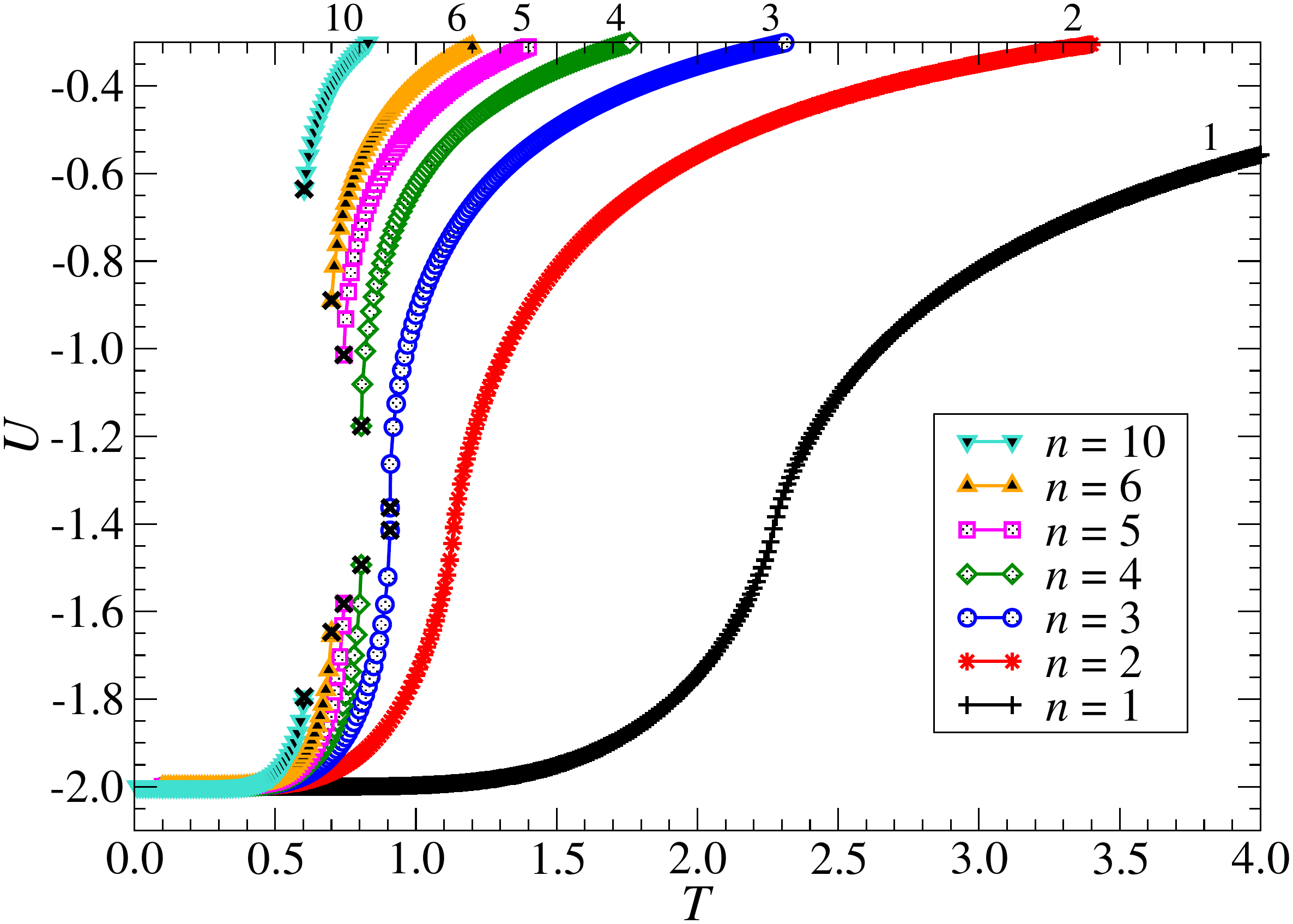}
\caption{Temperature dependence of the internal energy $U$ for various $n$.
The presence of the discontinuity in $U$ at the 
cross-over temperature 
$T_n^{*}$ for $n \geq 3$ is emphasized by the symbols $\times$.}
\label{fig4}
\end{figure}

We have recognized that $L_4^{~}$ is larger than $L_3^{~}$. 
In order to roughly capture the $n$-dependence in the latent heat $L_n^{~}$, 
we also calculate the cases with the higher internal dimensions ($n \ge 5$).
Figure~\ref{fig4} shows the internal energy $U( T )$ for those cases from $n = 1$ to
$n = 10$. For the cases $n = 5$, $6$, and $10$, respectively, the value of $m$ is
chosen to be $m = 120$, $50$, and $50$. Discontinuous nature in $U( T )$ 
is evident for $n \ge 3$. The observed discontinuity $L_n^*$ at each cross-over temperature 
$T_n^{*}$ is summarized in Tab.~\ref{tab1}. 
\tabcolsep=6pt
\begin{table}[tb] 
\caption{The list of the cross-over temperature and the jump in the calculated internal energy
for higher $n$. The value of $m$ used in the CTMRG calculation is also shown.}
\begin{center}
\begin{tabular}{|c|c|c|c|} \hline
$n$ & $m$ & 
$T_n^{*}$ & 
$L^*_n$ \\ \hline \hline
	$\phantom{0}5$ & $120$ & $0.74388(2)\phantom{0}$ & $0.568(7)\phantom{0}$ \\ \hline
	$\phantom{0}6$ & $\phantom{0}50$ & $0.70049(40)$ & $0.757(12)$ \\ \hline
	$10$ & $\phantom{0}50$ & $0.60326(5)\phantom{0}$ & $1.159(2)\phantom{0}$ \\ \hline
\end{tabular}
\end{center}
\label{tab1}
\end{table}
\tabcolsep=3pt
Although careful extrapolation with respect to $m$ is not
performed here, the increasing nature of the jump $L_n^*$ with 
respect to the internal dimension $n$ is apparent.
It can be conjectured that the latent heat $L_n^{~}$ is a monotonously
increasing function of $n$.

\section{Conclusions}

We have studied the thermodynamic properties of the cross-polytope models on the square 
lattice. We have focused on the two cases in which the internal dimensions of the site
vectors were $n = 3$ and $n = 4$. The free energy and the internal energy were calculated
by means of the CTMRG method. The presence of the first-order phase transition is confirmed
for both models from the temperature dependence of these functions.
For the 16-cell model ($n = 4$), we evaluated the latent heat $L_4^{~} = 0.3172$, which is
larger than that for the octahedron model $L_3^{~} = 0.0516$. The increasing tendency
in the latent heat with respect to $n$ is similar to the latent heat of the $q$-state
Potts models, which is increasing with $q$ when $q \geq 5$.

It is worth mentioning that the octahedron model ($n = 3$) is similar to the $5$-state
Potts model, in the point that there are $4$ type of single spin flip, which increases the
energy by $4 J$ on the square lattice, from the completely ordered ferromagnetic ground
state. This is the reason why both the octahedron and the 5-state Potts models reveal
the small latent heat. Similar correspondence can be considered between the 16-cell model
($n = 4$) and $7$-state Potts model.

In four dimensions, there are exceptional polytope models, which are $24$-, $120$-,
and $600$-cell models. There is an interest in the clarification of the nature of 
their phase transitions.  It should be noted that these models have rich sub-group
structure in their site-vector symmetry. Since these models contain huge amount of
the site degrees of freedom, algorithmic improvements of the CTMRG method is necessary
in order to carry out their numerical investigation.

\begin{acknowledgments}

This work was funded by Agent\'{u}ra pre Podporu V\'{y}skumu a V\'{y}voja (APVV-16-0186 EXSES) and Vedeck\'{a} Grantov\'{a} Agent\'{u}ra M\v{S}VVa\v{S} SR a SAV (VEGA Grant No. 2/0123/19). T.N. and A.G. acknowledge the support of Ministry of Education, Culture, Sports, Science and Technology (Grant-in-Aid for Scientific Research JSPS KAKENHI 17K05578 and 17F17750). This publication was made possible through the support of the ID\#~61466 grant from the John Templeton Foundation, as part of the "The Quantum Information Structure of Spacetime (QISS)" Project (qiss.fr). The opinions expressed in this publication are those of the author(s) and do not necessarily reflect the views of the John Templeton Foundation.
\end{acknowledgments}

\end{document}